\documentclass[useAMS,usenatbib,letters]{mn2e}
\usepackage{times,graphicx,amsmath,amsfonts,amssymb,aas_macros}

\newcommand{\ztwo}{$z\sim2$}
\newcommand{\zzero}{$z=0$}

\newcommand{\afe}{[$\alpha$/Fe]}

\newcommand{\msun}{M$_{\odot}$}                            % Msun
\newcommand{\mstar}{M$_{*}$}                                % Mstar
\newcommand{\msunyr}{\msun\ yr$^{-1}$}              % Msun / yr
\newcommand{\vsigma}{$v/\sigma$}

\newcommand{\xx}{gas-rich clump-driven}

\title[Formation of Metal-Rich Globular Clusters]{Star-Forming Galaxies at {\boldmath \ztwo}\ and the Formation of the Metal-Rich Globular Cluster Population}

\author[K. L. Shapiro et al.]{Kristen L. Shapiro,$^{1}$
Reinhard Genzel,$^{2,3}$ Natascha M. F\"orster Schreiber$^{3}$ \\
$^{1}$UC Berkeley Department of Astronomy, Berkeley, CA 94720, USA \\
$^{2}$UC Berkeley Department of Physics, Berkeley, CA 94720, USA \\
$^{3}$Max-Planck-Institut f\"ur extraterrestrische Physik (MPE), Giessenbachstr. 1, 85748 Garching, Germany
}

\begin{document}

\date{Accepted	2009 December 23. Received 2009 November 20; in original form 2009 September 8}

\pagerange{\pageref{firstpage}--\pageref{lastpage}} \pubyear{}

\maketitle

\label{firstpage}

%%%%%%%%%%%%%%%%%%%%%%%%%%%%%%%%%%%%%%%%%%%%%%%
% ABSTRACT

\begin{abstract}
We examine whether the super star-forming clumps ($R\sim1-3$ kpc; ${\rm M}\sim10^8-10^{9.5}$ \msun) now known to be a key component of star-forming galaxies at \ztwo\ could be the formation sites of the locally observed old globular cluster population.  We find that the stellar populations of these super star-forming clumps are excellent matches to those of local metal-rich globular clusters.  Moreover, this globular cluster population is known to be associated with the bulges / thick disks of galaxies, and we show that its spatial distribution and kinematics are consistent with the current understanding of the assembly of bulges and thick disks from super star-forming clumps at high redshift.  Finally, with the assumption that star formation in these clumps proceeds as a scaled-up version of local star formation in molecular clouds, this formation scenario reproduces the observed numbers and mass spectra of metal-rich globular clusters.  The resulting link between the turbulent and clumpy disks observed in high-redshift galaxies and a local globular cluster population provides a plausible co-evolutionary scenario for several of the major components of a galaxy: the bulge, the thick disk, and one of the globular cluster populations.
\end{abstract}

%%%%%%%%%%%%%%%%%%%%%%%%%%%%%%%%%%%%%%%%%%%%%%%
% SUBJECT HEADINGS


%%%%%%%%%%%%%%%%%%%%%%%%%%%%%%%%%%%%%%%%%%%%%%%
% ACTUAL PAPER

%%%%%%%%%%%%%%%%% INTRODUCTION %%%%%%%%%%%%%%%%%%%%%

\section{Introduction}
\label{sec:intro}

The Universe at \ztwo\ is now known to be an important epoch in galaxy formation, during which the cosmic star formation rate density peaks as galaxies undergo rapid growth \citep[e.g.][]{Rud+03}.  Much of this activity occurs in massive, rapidly star-forming galaxies (\mstar~$\sim10^{10}-10^{11}$~\msun, SFR~$\sim10-200$~\msunyr; \citealt{Red+05}) identified via their rest-frame optical and near-infrared colors.  Comparisons of the observed properties of this population (clustering, dynamical masses, SFRs) with dark matter halo properties in cosmological simulations imply that these galaxies will evolve into local bulge-dominated spiral, lenticular, and low-mass elliptical galaxies, increasing in halo mass by a factor of three between \ztwo\ and \zzero\ \citep{Con+08,Genel+08}.  For the majority of the population, this evolution will occur in a ``smooth" fashion, via accretion and minor mergers, with an average of $0-1$ major mergers during this interval \citep{Genel+08}.  This smooth yet rapid mass growth, coupled with the lack of disruption by major mergers, renders these galaxies a natural population in which important structures in local galaxies may be assembled and thus a critical population to study.
 
Detailed dynamical and morphological observations of this galaxy population have revealed that a substantial fraction is characterized by large ($\sim5-10$~kpc), regularly rotating, thick disks ($h_z\sim1$~kpc, $v/\sigma\sim2-6$; \citealt{For+06,For+09,Cre+09,ElmElm05,Elm+09}).  These galaxies are each populated by $5-10$ super star-forming (super-SF) clumps ($R\sim1-3$~kpc), which collectively account for $\sim30$\%\ of the total baryonic mass in each galaxy \citep{CowHuSon95,vdBer+96,ElmElm05,Gen+08,Elm+09}.  The masses of individual super-SF clumps are limited at the upper end by the ``Toomre" mass, the characteristic scale of these marginally stable ($Q\sim1$) rotating galaxies, ${\rm M}_T\sim2.5\times10^9$~\msun, and are believed to have typical masses ${\rm M}\sim10^9$~\msun\ \citep{Gen+08,Elm+09,DekSarCev09}.  Simulations suggest that these clumps form naturally in gas-rich turbulent disks and that dynamical friction will cause them to spiral in to the center of the host galaxy on timescales $\lesssim1$~Gyr and form a nascent bulge \citep{Nog99,Imm+04,BouElmElm07}, leaving a fraction of their mass behind in a thin star-forming disk and a quiescent thick disk \citep{BouElmMar09}.  Similar conclusions are reached from observations of galaxies in this ``\xx\ phase" \citep{Gen+08,ElmBouElm08a}, which we will refer to here as \ztwo\ star-forming galaxies (z2SFGs).

Could this phase of galaxy evolution also be associated with the formation of globular clusters (GCs)?  The high masses (\mstar~$\sim10^4-10^6$~\msun) and densities ($\rho_{central}\sim8\times10^3$~\msun~pc$^{-3}$) of GCs require exceptionally massive and/or dense giant molecular clouds (GMCs); the super-SF clumps found in z2SFGs are thus good candidates for this process.  Indeed the average properties of the super-SF clumps ($R\sim1-3$~kpc; M~$\sim10^9$~\msun) are very consistent with the predictions of \citet{HarPud94} and \citet{McLauPud96}, who concluded that ``super GMCs" in the early Universe ($R\sim1$~kpc; M~$\sim10^9$~\msun) are likely formation sites for GCs (see \S\ref{sec:discuss} for a discussion of other proposed GC formation mechanisms).  Subsequently, \citet{ElmEfr97} showed that such super~GMCs~/~super-SF~clumps are naturally formed in an interstellar medium characterized by high turbulent velocities and gas surface densities.  Likewise, \citet{EscLar08} showed that gas-rich disk galaxies would have large Jeans masses and could create the requisite massive super GMCs.  These authors mention in passing that such disks may correspond to the turbulent and gas-rich galaxies observed at high redshift.

In this letter, we expand this idea and propose that the observed features of~\zzero\ GC populations can be explained by assuming they formed in super-SF clumps in z2SFGs.  To test this hypothesis, we investigate whether z2SFGs can account for the three main characteristics of local GC systems: 
\begin{itemize}
\item the stellar populations of GCs, 
\item the spatial distributions and kinematics of GC systems, and 
\item the numbers and mass spectra of GCs within galaxies.  
\end{itemize}
Globular clusters are a bimodal population in metallicity; metal-poor GCs increase in number with galaxy mass, and metal-rich GCs with bulge mass.  Detailed observations of the Milky Way GC population additionally show that metal-poor GCs are associated spatially and kinematically with the galaxy halo and metal-rich GCs with the bulge and thick disk.  Comparing these properties to those of the super-SF clumps in z2SFGs (\S\ref{sec:compare}), we find that z2SFGs are plausible formation sites for metal-rich GCs.  This connection between an observed galaxy population and a globular cluster population provides further insight into the formation of both metal-rich and metal-poor GCs (\S\ref{sec:discuss}).

Throughout this letter, we assume a $\Lambda$-dominated cosmology with $H_{\rm 0}=70$~km~s$^{-1}$~Mpc$^{-1}$, $\Omega_{\rm m}=0.3$, and $\Omega_{\rm \Lambda}=0.7$.

%%%%%%%%%%%%%%%%% STELLAR POPS %%%%%%%%%%%%%%%%%%%%

\section{Comparison of Globular Clusters and {\boldmath \ztwo}\ Star-Forming Galaxies}
\label{sec:compare}

\subsection{Stellar Populations}
\label{sec:pops}

Stellar population ages, chemical abundances, and metallicities provide important constrains on the formation epoch, duration, and sites of GCs.  The ages of globular clusters are measured via the location of their stars in color-magnitude diagrams (for Milky Way GCs) and by comparing stellar absorption line indices in their spectra to single stellar population models (for GCs in other galaxies).  However, both methods are limited by current observational and theoretical understanding of evolved stellar populations and can thus only constrain absolute ages to within a few Gyr for the oldest populations (age~$\gtrsim10$~Gyr).  Current estimates of the ages of GCs are $9-12$~Gyr in the Milky Way \citep[e.g.][]{deAng+05,MenProFor09} and $10\pm2$~Gyr in other galaxies \citep[e.g.][]{PuzKisGou06}.  Such ages are comparable to the look-back time at \ztwo\ of $10.2$~Gyr but are consistent with formation redshifts $z\sim1.5-4$.

The formation timescale for GCs is known, via their high observed $\alpha$-enhancement (\afe~$\sim0.3$), to be rapid enough that Type Ia supernovae have not yet exploded and enriched the interstellar medium with Fe-peak elements ($0.7-1$~Gyr; \citealt{ScaBil05}).  GC formation must therefore occur during a relatively brief epoch of high SFR and significant mass growth; this is indeed the case for the \xx\ phase in z2SFGs, which has a duty cycle of $0.5-1$~Gyr \citep{Gen+06,Gen+08,Dad+07a}.  The duration of this phase is limited by dynamical friction on the clumps, which causes them to lose angular momentum and migrate from the disk to the galaxy center \citep{Nog99,Imm+04,BouElmElm07,Gen+08}, as well as by stellar feedback, which may disrupt the molecular component of individual clumps even earlier \citep{MurQuaTho09}.  The $\alpha$-element enhancement from this short and powerful star formation event has been directly observed in z2SFGs and higher-redshift analogs to be $0.25-0.7$~dex \citep{Pet+02,Hal+08,Qui+09}, comparable to those of GCs.

\begin{figure}
	\centering
	\includegraphics[width=7.5cm,trim=1.3cm 0.5cm 0cm 1cm,clip]{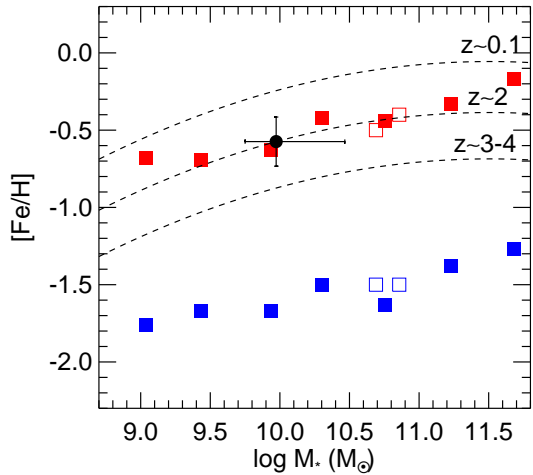}
	\caption{The observed relation between the host galaxy mass and the mean metallicities of the metal-rich ({\it filled red squares}) and metal-poor ({\it filled blue squares}) globular cluster populations for early-type Virgo Cluster galaxies \citep{ACSVCSXI}.  The globular cluster populations of the Milky Way and M31 are also included ({\it open squares}).  \citet{Hal+08} have measured stellar [Fe/H] metallicities for a coadded average spectrum of z2SFGs ({\it black circle}); here we expand the horizontal error bar to include the average factor three mass growth between \ztwo\ and \zzero\ (see \S\ref{sec:number}).  To compare with star-forming galaxies at other redshifts, the approximate locations of the galaxy mass-metallicity relationships are indicated ({\it dashed lines}).  We use the \ztwo\ mass-metallicity relation of \citet{Erb+06a} and assume \afe~$\sim0.3$ (see text) to convert [O/H] to [Fe/H] (assuming [O/H]~$\approx$~[Z/H]~$=$~[Fe/H]$+ 0.94$\afe; \citealt{ThoMarBen03}).  \citet{Erb+06a} and \citet{Man+09} show that the $z\sim0.1$ (SDSS) and $z\sim3-4$ (for Lyman-break galaxies) relations are $\sim0.3$~dex above and below that at \ztwo, respectively, and the approximate locations of these relations are indicated.}
	\label{fig:metal}
\end{figure}

Finally, the most stringent constraint on GC formation comes from the metallicities of these systems, which have a clearly bimodal distribution.  In most galaxies, the peaks of this distribution are near [Fe/H]~$=-1.5$~and~$-0.5$, with a slight dependence on the mass of the host galaxy \citep[e.g.][]{BroStr06,ACSVCSXI}.  In Figure~\ref{fig:metal}, we compare this relationship to an average measurement of [Fe/H] in z2SFGs \citep{Hal+08}.  For additional reference, the approximate locations of the mass-metallicity relationships at several redshifts are indicated.  The metal-poor GCs have metallicities far below those observed in z2SFGs and therefore almost certainly did not form during the \xx\ phase observed in z2SFGs.  In contrast, the metal-rich GCs overlap impressively with the galaxy masses and metallicities of z2SFGs.

This agreement of the look-back time, star formation duty cycle, and metallicity of z2SFGs with the ages, elemental abundances, and metallicities of metal-rich GCs suggests that this subpopulation of GCs could have been formed in z2SFGs.

%%%%%%%%%%%%%%% DISTRIBUTION AND KINEMATICS %%%%%%%%%%%%%%%%%

\subsection{Spatial Distribution and Kinematics}
\label{sec:kinem}

The formation scenario for GCs is also strongly constrained by their phase space distribution, which has been well measured only in the Milky Way.  GC systems similar to that of the Galaxy have been observed in other local spiral galaxies (M31:\citealt{HucBroKen91,Bar+00} and M81:\citealt{Sch+02}) but have not yet been confirmed with large samples of GCs in lenticular and low-mass elliptical galaxies \citep[][but see e.g. \citealt{Bri+97,Kun+02}]{BroStr06}.  In this section, we therefore must assume that the spatial distributions and dynamics of GC populations in the Milky Way are representative of those in bulge-dominated spiral, lenticular, and low-mass elliptical galaxies; this allows us to examine whether the metal-rich GCs in such local galaxies could have been produced in z2SFGs.

The Milky Way's metal-rich GC population is associated with both the bulge and the thick disk (Table~\ref{tab:comp}).  These GCs are centrally concentrated, with the majority of the population distributed spherically in the Galaxy center (i.e. de Vaucouleurs profile) and having bulge-like kinematics \citep{Cot99}.  A subset ($\sim1/3$) of the metal-rich GCs are found at larger radii ($>4$ kpc) and are rotationally supported (\vsigma~$\gtrsim1$; \citealt{Cot99}).  Based on the similar kinematics and stellar populations of these  ``disk" GCs and old thick disk stars, \citet{Zin85} proposed that these GCs were formed during a ``transient, thick-disk phase."  This prescient hypothesis virtually predicted modern observations of the \xx\ phase in z2SFGs ($h_z\sim1$~kpc; \vsigma~$\sim2-6$).

The current understanding of this phase further elucidates the connection between the metal-rich GCs potentially formed in super-SF clumps, thick disks, and bulges.  As the clumps (initially at $r\sim5-10$~kpc) migrate towards the center of their host z2SFG to form a proto-bulge ($r\lesssim3$~kpc; \citealt{Gen+08}), they presumably transport most of their newly-formed GCs with them, with a smaller fraction of the GCs being stripped off during this process and remaining in the galaxy disk.  This behavior has been seen in simulations by \citealt{ElmBouElm08b}, who populate each clump with a point mass of $\sim10^5-10^6$~\msun\ (used in their model to represent intermediate mass black holes) and find that most, but not all, of these point masses migrate to the galaxy center with their host clumps on short timescales ($t\lesssim1$~Gyr).  The final GC distribution and kinematics in fact mimic those of the stars, the majority of which are carried into the central few kpc, leaving behind a fraction of the mass in the (thick) disk \citep{BouElmMar09}.  Those GCs that migrate all the way to the galaxy center will interact with other GCs, other clumps, and the assembling bulge; this scattering will cause them to lose some of their rotational support, creating the low \vsigma\ ``bulge" GC population (Table~\ref{tab:comp}).  Some of these bulge GCs may be tidally disrupted and destroyed during this process, and some may sink to the galaxy center due to dynamical friction (see also \S\ref{sec:number}).  In contrast, the GCs that remained in the thick disk would not interact with other GCs or clumps and would therefore maintain the rotational support of the original clumpy thick disk (\vsigma~$\sim2-3$ in thick-disk GCs; \vsigma~$\sim2-6$ in z2SFGs).

The GC system that would result from a z2SFG would therefore be expected to have a primary component associated with the bulge and a less populous component associated with the thick disk, in excellent agreement with the observed properties of the metal-rich GC population.  This scenario additionally explains why this GC population is associated with both the bulge and the thick disk.

\begin{table}
\caption{Distribution and Kinematics of Galactic Components and z2SFGs}
\label{tab:comp}
\begin{center}
\begin{tabular}{lccc}
%\begin{tabular}{lccccc}
\hline
 				& Profile 		& Extent (kpc)		& \vsigma 		\\
%				& Type 		&  (kpc)			&  			\\
% 				& Profile 		& Extent 			& \vsigma 		\\
% 				& Type 		&  (kpc)			&  			\\
\hline
\hline
z2SFGs 			& Disk		& $h_z\sim1$		& $2-6$ \\
 & & &  \\
Thick Disk Stars 	& Disk		& $h_z\sim0.8-1.4$	& $2-4$ \\
MR (Disk) GCs	 	& Disk		& $h_z\sim0.8-1.5$	& $2-3$ \\
Bulge Stars 		& de Vaucouleurs	& $R_e\sim1.2$	& $1-1.5$ \\
MR (Bulge) GCs	 & de Vaucouleurs	& $R_e<1.7$	& $0.4-0.6$ \\
 & & & \\
Halo Stars 		& $R^{-3.5}$	& $R<40$ 		& $0.2-0.8$ \\
MP (Halo) GCs 	& $R^{-3.5}$	& $R<40$			& $0.3-0.4$ \\
 & & & \\
Thin Disk Stars 	& Disk		& $h_z\sim0.3$		& $6-12$ \\
% 				& Age 		& \afe 		& \feh 		& Distribution \\
% 				& (Gyr) 		&  			&  			&  (kpc) \\
%\hline
%\hline
%z2SFGs 			& \ 10.2$^b$	& 0.25 -- 0.7 	& -1.0 -- -0.4	& Disk, $h_z\sim1$ \\
% & & & &  \\
%Thick Disk Stars 	& $\gtrsim$10	& 0 -- 0.4		& -1.6 -- -0.3	& Disk, $h_z\sim1.4$ \\
%MR (Disk) GCs$^a$ 	& 9 -- 12		& $\sim$0.3 	& -1.0 -- -0.2	& Disk, $h_z\sim1.1$ \\
%Bulge Stars 		& $\gtrsim$10	& 0 -- 0.4 		& -1.5 -- +0.5	& deV$^c$, $R_e=1.2$ \\
%MR (Bulge) GCs$^a$ & 9 --12		& $\sim$0.3	& -0.8 -- -0.2	& deV$^c$, $\bar{R}\sim1.7$ \\
 %& & & & \\
%Halo Stars 		& $>$10		& 0.2 -- 0.5 	& -4.5 -- -0.5	& $R^{-3.5}, R\sim4-50$ \\
%MP (Halo) GCs$^a$ 	& 9 -- 12		& $\sim$ 0.3	& -2.3 -- -0.8	& $R^{-3.5}, R\sim4-40$ \\
% & & & & \\
%Thin Disk Stars 	& 0 -- 12		& -0.2 -- 0.2 	& -0.5 -- +0.3	& Disk, $h_z\sim0.3$ \\

% 				& Age 		& \afe 		& \feh 		& \vsigma 		& $h_z$ \\
%z2SFGs 			& \ 10.2$^b$	& 0.25 -- 0.7 	& -1.0 -- -0.4	& 2 -- 6		& $\sim$1 \\
% & & & & & \\
%Thick Disk Stars 	& $\gtrsim$10	& 0 -- 0.4		& -1.6 -- -0.3	& 2 -- 4		& 0.8 -- 1.4 \\
%MR (Disk) GCs$^a$ 	& 9 -- 12		& $\sim$0.3 	& -1.0 -- -0.2	& 2 -- 3		& 0.8 -- 1.5 \\
%Bulge Stars 		& $\gtrsim$10	& 0 -- 0.4 		& -1.5 -- +0.5	& 1 -- 1.5		& \ 1.2$^c$ \\
%MR (Bulge) GCs$^a$ & 9 --12		& $\sim$0.3	& -0.8 -- -0.2	& 0.4 -- 0.6	& $\sim$1.7$ \\
% & & & & & \\
%Halo Stars 		& $>$10		& 0.2 -- 0.5 	& -4.5 -- -0.5	& 0.2 -- 0.8	& 2.7$^c$ \\
%MP (Halo) GCs$^a$ 	& 9 -- 12		& $\sim$ 0.3	& -2.3 -- -0.8	& 0.3	 -- 0.4	& $\sim$5$^d$ \\
% & & & & & \\
%Thin Disk Stars 	& 0 -- 12		& -0.2 -- 0.2 	& -0.5 -- +0.3	& 6 -- 12		& 0.3 \\

\hline
\end{tabular}
\end{center}
Comparison of metal-rich (MR) and metal-poor (MP) GC populations to components of the Milky Way and to z2SFGs (see definition in \S\ref{sec:intro}).  Although the descendants of z2SFGs are likely more bulge-dominated than the Milky Way, the spatial and kinematic properties of GCs in these galaxies are not yet well constrained; here we assume the GC population of the Milky Way to be representative in order to evaluate the possible link with z2SFGs.  Data are from \citet{Arm89}, \citet{Cot99}, \citet{Cre+09}, \citet{ElmElm06}, \citet{Min95}, and \citet{Zin85}.
%$^a$ MR = metal-rich; MP = metal-poor \\
%$^b$ Age of Universe at observed redshift (\ztwo) \\
%$^c$ de Vaucouleurs profile \\
%$^d$ Half-light radius of $r^{-3}$ profile extending to 40 kpc \\
%The comparison shown here is for the Milky Way.  Stellar populations in the Milky Way stars and GCs are observed to be similar to those of the bulge-dominated galaxies that are the likely descendants of z2SFGs (\S\ref{sec:pops}).  The kinematic properties of GC populations in these galaxies is not yet well constrained, so the Milky Way is used here as an example (\S\ref{sec:kinem}).  Data are quoted from \citet{Arm89}, \citet{Cot99}, \citet{Cre+09}, \citet{deAng+05}, \citet{ElmElm06}, \citet{Hol+93}, \citet{MenProFor09}, \citet{Min95}, \citet{ACSVCSXI}, \citet{PuzKisGou06}, \citet{Zin85}, \citet{Zoc+03}, and references therein.
\end{table}
% Fe/H, h_z for Thin Disk, Thick Disk, Bulge, Halo taken from C&O 1996.
% a/Fe = 1/4 (Mg/Fe + Ca/Fe + Si/Fe + Ti/Fe)

%%%%%%%%%%%%%%%%% NUMBERS AND MASSES %%%%%%%%%%%%%%%%%%%%

\subsection{Number and Mass Function}
\label{sec:number}

To evaluate whether super-SF clumps in z2SFGs can produce the number and mass distribution of the relevant (i.e. metal-rich) GC population, we employ a simple analytic model.  We begin by estimating the number of metal-rich GCs expected in the \zzero\ descendants of z2SFGs.  In the local Universe, the parameter $T=N$~$/$~(M$_{gal}/10^9$~\msun) is used to express the number of GCs per $10^9$ \msun\ of galaxy stellar mass \citep{ZepAsh93,RhoZepSan05}.  \citet{ACSVCSXV} have measured this quantity for a large range of galaxy masses in Virgo early-type galaxies; they find a nearly constant value of $T = 5$ for galaxies with \mstar~$=0.02-2\times10^{11}$ \msun.  At \ztwo, the z2SFGs have stellar masses of $\sim10^{10}-10^{11}$~\msun\ \citep[e.g.][]{For+09}.  Assuming the factor three increase in halo mass expected between \ztwo\ and \zzero\ \citep{Con+08,Genel+08} includes a constant baryon fraction in the accreted material, this implies stellar masses at \zzero\ of $3\times10^{10}-3\times10^{11}$ \msun\ and thus $T=5$ (or $\sim150-1500$ GCs) is appropriate for this population.  (Similar predictions of GC numbers are also obtained using the universal cluster formation efficiency of \citealt{McLau99}.)  In this galaxy mass range, $20-40$\%\ of GCs are observed to be metal-rich \citep{ACSVCSIX,ACSVCSXV}, implying that z2SFGs should produce roughly $30-600$ metal-rich GCs that persist to \zzero.

We now evaluate whether the super-SF clumps in z2SFGs can produce the requisite number of metal-rich GCs during the \xx\ phase.  Star formation within super-SF clumps has been explored analytically \citep{HarPud94,McLauPud96} and has been observed for the GC-analogs, young massive clusters (YMCs), which form locally in gas-rich (e.g. LMC, M82) and high SF (e.g. the Antennae) environments \citep{ElsFal85,ElmEfr97,EngFre03,deGriBasLam03,Wil+03}.  These studies have concluded that star formation in super-SF regions plausibly proceeds in much the same fashion as in their less massive cousins, local GMCs; super-SF regions form gravitationally unstable structures with an efficiency of a few percent and with a power law mass spectrum of index $-2<\alpha<-1.5$, the latter being a result of the turbulent cascade.  The mass spectrum is limited at the high end, as in local GMCs, at $\sim10^{-3}$ times the mass of the cloud, so super-SF clumps ($\sim10^9$~\msun) form structures with masses up to $10^6$~\msun\ (i.e. YMCs, potential GCs), while local GMCs form structures with masses up to $10^3$~\msun\ (i.e. open clusters).  At the low end of the mass spectrum, less is known about the limiting mass; here, we assume a similar scaling with cloud mass, such that the lower mass limit in local GMCs of $10^{-6}$ times the cloud mass ($\sim1$~\msun) applies also to super-SF clumps.

\begin{figure}
	\centering
	\includegraphics[width=7.5cm,trim=1cm 0.3cm 0cm 1cm,clip]{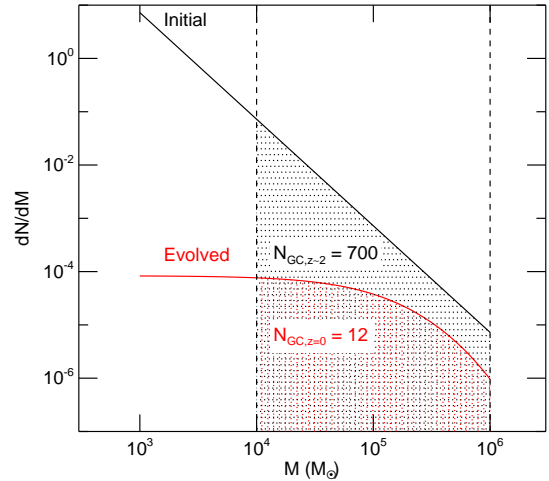}
	\caption{Predicted mass spectrum for structures formed in a z2SFG super-SF clump.  The slope of the initial mass spectrum (\ztwo; {\it black}) is observed to be common among all structures in the ISM (clouds, cores, open clusters, YMCs).  The upper and lower limits are set by assuming that a molecular cloud forms structures with $10^{-6}-10^{-3}$ of its mass, as observed in local GMCs, and the normalization is set such that $\sim5$\%\ of the original cloud mass will form stars, as observed locally \citep[e.g.][]{Wil+03}.  The evolutionary model of \citet{ACSVCSXII} is applied to the initial mass spectrum and results in a preferential decrease of low-mass structures in the evolved mass spectrum (\zzero; {\it red}).  Vertical dashed lines indicate the mass range in which bound structures are observed as GCs, and the shading and labels indicate the number of GCs produced per clump in this model.}
	\label{fig:struct}
\end{figure}

Combining this information with observations of super-SF clumps in z2SFGs, we predict the number and mass function of (metal-rich) GCs that can be produced.  Letting each super-SF clump form stellar clusters according to the mass range and spectrum described above results in $\sim700$ objects per clump with mass $10^4-10^6$~\msun, distributed with a power law of assumed slope $\alpha=2$ (Figure~\ref{fig:struct}).  \citet{FalZha01} and \citet{ACSVCSXII} have shown that local GC mass spectra are not power laws but can instead be well-represented by ``evolved" Schechter functions that flatten below a characteristic turn-over mass ($\sim2\times10^5$~\msun), which they suggest results from the evaporation of lower mass GCs through two-body relaxation \citep[see also][]{McLauFal08}.  At the high mass end (above a cutoff mass M$_c$), the distribution drops more steeply than a power law; this may be the result of higher mass GCs preferentially sinking to the galaxy center with their host super-SF clumps through dynamical friction (\S\ref{sec:kinem}) or being destroyed by gravitational shock heating in the potential of the host galaxy \citep{FalZha01,McLauFal08}.  \citet{ACSVCSXII} show that the evolved and initial mass functions can be straightforwardly linked (their equations 4-7), with the exact shape of the final mass function dependent on only two parameters, a cutoff mass (M$_c$) and a cumulative mass loss per GC ($\Delta$).  Here, we adopt their values for M$_c$ and $\Delta$ appropriate for galaxies with masses characteristic of z2SFG descendants.  Integrating the resulting evolved GC mass function suggests that, of the initial $\sim700$ GCs per clump, $\sim12$ will survive from \ztwo\ to \zzero\ (Figure~\ref{fig:struct}).  With $5-10$ clumps per galaxy, we therefore expect $3500-7000$ GCs to be created during a galaxy's \xx\ phase, of which $\sim60-120$ survive to \zzero.

This order-of-magnitude calculation shows that z2SFGs may plausibly produce metal-rich GCs whose numbers and mass spectra compare well with GCs observed in local bulge-dominated galaxies.  Variations in our simple assumptions of super-SF clump masses and numbers would increase the scatter in the predicted number of metal-rich GCs ($60-120$) to more closely match the full range observed locally in likely z2SFG descendants ($30-600$).

%%%%%%%%%%%%%%%%% DISCUSSION %%%%%%%%%%%%%%%%%%%%

\section{Discussion and Conclusions}
\label{sec:discuss}

In the literature, two broad models of GC formation are discussed: 1) Each galaxy halo creates a single GC population of roughly uniform metallicity, with the second GC population being a product of hierarchical merging (metal-poor GCs accreted from dwarf galaxies: \citealt{CotMarWes98}; metal-rich GCs created during gas-rich mergers: \citealt{AshZep92}).  2) Each galaxy halo has two distinct episodes of GC formation, with metal-poor GCs forming first during the collapse of the protogalactic cloud and metal-rich GCs in a subsequent phase of star formation (``in situ": \citealt{ForBroGri97}).  On the basis of the relation between galaxy luminosity (mass) and GC metallicity for both GC subpopulations (Figure~\ref{fig:metal}), \citet{StrBroFor04} have argued that the ``in situ" model is strongly preferred, since the accretion or creation of GCs during mergers would blur correlations between the properties of galaxies and their GCs.

Our hypothesis augments this model by providing a natural mechanism for the coincident formation of metal-rich GCs, bulges, and thick disks.  In \S\ref{sec:compare}, we have shown that the stellar populations, spatial distribution, kinematics, numbers, and mass spectra of metal-rich GCs are all consistent with formation in super-SF clumps during the \xx\ phase of a galaxy's evolution.  The rapid ($\lesssim1$~Gyr) migration of these clumps to the centers of their host galaxies provides an effective means of transporting the majority of GCs to the vicinity of the proto-bulge formed from the coalescing clumps; this same mechanism also accounts for the thick-disk GCs, which, like the newly formed thick-disk stars, are stripped from their host clumps and left behind at large radii.  This process has a built-in truncation mechanism: once a bulge has formed, it stabilizes the remaining disk against further fragmentation into large clumps \citep{BouElmElm07}.  Since the clumps are responsible for simultaneous GC, bulge, and thick disk formation, these must occur as a single, short ($\lesssim1$~Gyr), and very significant event in the lifetime of a galaxy.

This scenario provides new insight into the process of galaxy assembly at high redshift.  The formation of metal-rich GCs in the \xx\ phase implies that the number of these GCs in a galaxy will be closely tied to the importance of this phase in the galaxy's history.  Galaxies with a more dramatic \xx\ phase will produce more metal-rich GCs, and the coalescing clumps will create larger bulges, resulting in the observed correlation between bulge mass and the number of metal-rich GCs \citep{ACSVCSIX}.  The potential association of galaxies at the massive end of this correlation (e.g. M87) with quiescent spheroids at \ztwo\ \citep[e.g.][]{Tru+06b,Tac+08} suggests that the populous metal-rich GC populations in these local giant ellipticals may result from a \xx\ phase that was very dramatic and occurred prior to \ztwo\ during the galaxies' primary SF epochs or during a merger of two galaxies in the \xx\ phase \citep[see also e.g.][]{RhoZep04}.  Conversely, the absence of significant metal-rich GC populations (and bulges) in late-type galaxies suggests that massive, migrating super-SF clumps did not play an important role in late-type galaxy formation.

Finally, the scenario proposed here also constrains metal-poor GC formation through the current understanding of the formation history of z2SFGs.  Significant observational and theoretical evidence indicates that z2SFGs are assembled via the rapid and smooth transport of cold ($T<T_{vir}$) gas along filaments into the centers of halos \citep[e.g.][]{Gen+06,Dad+07a,Erb08,DekSarCev09}.  This inflow is almost certainly also present at the earlier epochs in which metal-poor GCs form.  These GCs may thus be formed within small concentrations inside filaments or at the intersection of filaments inside the collapsing dark matter halo.  This insight is essentially an update of the dissipational collapse model of \citet{SeaZin78} and \citet{HarPud94} with the detailed information now available at high redshift, and it provides a simple mechanism through which the formation of metal-poor GCs is linked to that of the halo.

%%%%%%%%%%%%%%%%%%%%%%%%%%%%%%%%%%%%%%%%%%%%%%%
% ACKNOWLEDGEMENTS

\section*{Acknowledgments}

We are grateful to the referee, Dean McLaughlin, whose comments significantly improved this paper.  We also wish to thank the SINS team, especially Andi Burkert, Ric Davies, Peter Johansson, Amiel Sternberg, and Linda Tacconi, for enlightening conversations and for comments on the manuscript.  This paper has benefitted from discussions with Harald Kuntschner, Chris McKee, Jeffrey Silverman, and Andrew Wetzel.  NMFS acknowledges support from the Minerva Program of the Max-Planck-Gesellschaft.

%%%%%%%%%%%%%%%%%%%%%%%%%%%%%%%%%%%%%%%%%%%%%%%
% BIBLIO

\bibliography{biblio}
\bsp

%%%%%%%%%%%%%%%%%%%%%%%%%%%%%%%%%%%%%%%%%%%%%%%
% END

\label{lastpage}

 \end{document}